RESEARCH PAPER

# Adaptive-Sliding Mode Trajectory Control of Robot Manipulators with Uncertainties

Mustafa M. Mustafa[1], Ibrahim Hamarash[1,2], Carl D. Crane[3]

[1]Department of Electrical Engineering, College of Engineering, Salahaddin University-Erbil, Kurdistan Region, Iraq
[2]Department of Computer Science and Engineering, University of Kurdistan Hewler, Iraq
[3]Department of Mechanical and Aerospace Engineering, University of Florida, FL, USA

**A B S T R A C T:**
In this paper, we propose and demonstrate an adaptive-sliding mode control for trajectory tracking control of robot manipulators subjected to uncertain dynamics, vibration disturbance, and payload variation disturbance. Throughout this work we seek a controller that is, robust to the uncertainty and disturbance, accurate, and implementable. To perform these requirements, we use a nonlinear Lyapunov-based approach for designing the controller and guaranteeing its stability. MATLAB-SIMULINK software is used to validate the approach and demonstrate the performance of the controller. Simulation results show that the derived controller is stable, robust to the disturbance and uncertainties, accurate, and implementable.

## 1. INTRODUCTION

Many robot manipulators in the modern world are required to work in high speed and high accuracy. For many applications, such as in the medical area, manufacturing printed circuit boards, robot manipulators are required to be manipulated accurately with high speed despite the uncertainty in the dynamics. Such a disturbance, that is caused by a vibration and/or a payload variation, causes a major uncertainty in the dynamics of robot manipulators. Without an efficient controller, that can reject this disturbance and deal with the uncertainty, the robot manipulator will perform poorly in terms of stability and accuracy. Therefore, the methods of robot manipulators control need to be developed to damp the disturbance torques while trajectory tracking control, specially the disturbance that is caused by vibration and payload variation.

Some position tracking control methods consider the disturbance as an uncertainty in the dynamics. While there are other methods consider this disturbance as exogenous torques to be damped separately. Therefore, many linear and nonlinear control approaches are used for trajectory tracking control. In (Economou *et al.*, 2000), the authors utilize finite impulse response filters to design a controller that suppress residual vibrations in flexible payloads carried by robot manipulators. However, the controller is restricted due to the limitations of these filters with nonlinear systems and the natural frequencies' knowledge requirements about the payload. In (Mamani, Becedas and Feliu, 2012), a sliding mode control (SMC) is used to control the position of a single link flexible robot arm. The controller rejects the disturbances that are caused by actuator friction and payload; however, the actuator bandwidth of this controller is beyond the classical limits. In (Feliu *et al.*, 2013), an innovative algorithm is designed to control a three degree of freedom flexible robot arm in the presence of vibrations. The control algorithm cancels the vibration and shows an accurate trajectory tracking control with acceptable control effort; however, the kinematics and the compliance matrix of the robot need to be known. In (Slotine and Li, 1987), another control algorithm is derived, based on the adaptive control theory, to solve the problem of trajectory tracking when the payload is unknown. The same theory is utilized by (Craig, Hsu and Sastry, 1987) to reject the exogenous disturbances while trajectory tracking, where the dynamics is assumed to be known. In (Hsia and Gao, 1990), the payload variation is treated within the dynamics of the robot by the proportional derivative (PD) control law. The authors of (Dawson *et al.*, 1990), use a PD controller with proper gains and initial conditions after bounding the disturbances, where they show uniformly bounded (UB) results of the tracking error. The input torques in (Craig, Hsu and Sastry, 1987; Slotine and Li, 1987) are better to that in (Dawson *et al.*, 1990; Hsia, Lasky and Guo, 1991; Spong, 1992) in term of implementation; however, these controllers are not robust to vibration and payload variation. Other works consider the fact that the disturbance cannot be linear parameterized, therefore, neural network and fuzzy logic methods are called through the control design (Dixon, Zergeroglu and Dawson, 2004; Gao *et al.*, 2018; Nafia *et al.*, 2018). However, there are many limitations with these controllers; therefore, the integral of the sign of the error (RISE) is utilized in (Cai, de Queiroz and Dawson, 2006; Patre *et al.*, 2006; Pedroza, MacKunis and Golubev, 2014; Shao *et al.*, 2018; Su, Xie and Li, 2019) to dominate the limitations and yield asymptotic tracking errors.

---
* **Corresponding Author:**
Mustafa M. Mustafa
E-mail: mustafa.atrushi@su.edu.krd
**Article History:**
Received: 22/12/2019
Accepted: 19/02/2020
Published: 08/09 /2020

The authors of (Mustafa, Hamarash and Crane, 2020), propose two nonlinear control approaches to control the position/displacement of robot manipulators in the presence of disturbance torques due to vibration and payload variation. In both proposed approaches the tracking error approaches zero or a region around zero. However, the estimation of the parameters is not considered while control design which decreases the control effort at the cost of accuracy. In this paper, we present an adaptive-sliding mode approach for trajectory tracking control of robot manipulators. The approach deals with the uncertainty in the dynamics and reject the disturbance of vibration and payload variation. The unknown parameters in the dynamics are estimated by using the adaptation law of the known parameters. Then, the disturbances are rejected separately by sliding mode control. The same disturbances that are formed for the first approach in (Mustafa, Hamarash and Crane, 2020) are called in this work to test the designed controller.

The paper is organized as follows: Section 2 introduces the dedicated dynamic model of the robot and preliminaries. In Section 3, the problem is formulated, and the controller is analyzed and designed. Simulation results are presented and analyzed is Section 4. The conclusions are given in Section 5.

## 2. THE DEDICATED DYNAMIC MODEL AND PRELIMINARIES

By considering uncertain disturbance torques caused by external vibration and payload variation, the dedicated dynamics for an n-degree-of-freedom, revolute-joints robot arm in joint-space coordinates is expressed as follows:

$$M(q)\ddot{q} + V_m(q,\dot{q})\dot{q} + G(q) = \tau - \tau_v - \tau_l \quad (1)$$

where $q$, $\dot{q}$, and $\ddot{q} \in \Re^n$ denote the angular displacement, angular velocity, and angular acceleration vectors, respectively; $M(q) \in \Re^{n \times n}$ represents the inertia matrix; $V_m \in \Re^{n \times n}$ represents the Coriolis-centripetal matrix; $G(q) \in \Re^n$ represents the gravity vector; $\tau \in \Re^n$ represents the torque input of the joints; $\tau_v \in \Re^n$ represents the disturbance due to external vibration; and $\tau_l \in \Re^n$ represents the payload variation torque.

For the subsequent controller design and analysis, we list the following property and assumption for the dynamics in Eq. (1) (Dixon *et al.*, 2013):

**Property 1.** The Coriolis and centrifugal matrix $V_m(q, \dot{q})$ can be appropriately determined such that

$$\chi^T \left( \dot{M}(q) - 2V_m(q, \dot{q}) \right) \chi = 0, \quad \forall \chi \in \Re^n \tag{2}$$

where $\dot{M}(q)$ is the derivative of the inertia matrix.

**Assumption 1.** Effects of the friction in the robot dynamics are neglected because it is out of the scope of this work.

## 3. PROBLEM FORMULATION AND CONTROL DESIGN

In this section the tracking control problem of robot manipulator is formulated in the presence of external vibration and payload variation. An auxiliary term is defined, and two assumptions are introduced regarding the properties of the position, velocity, and acceleration of the robot links. The auxiliary term and the assumptions are considered in the analysis and control design.

### 3.1. Problem Formulation

Robot manipulators are required to perform their tasks regardless the uncertainty in the dynamics and external disturbance, such as vibration and payload variation. Since the robot manipulator performs tasks by its end effector, we need to command the end-effector to a target point or following a desired trajectory within the workspace. An inverse kinematics process can be computed to obtain the joint-space trajectories (Crane III and Duffy, 2008; Hasan and Hamarash, 2017; Hasan, Crane III and Hamarash, 2019). For this manner, we need to design the input torque $\tau$ in Eq. (1), which drives the robot's links. In other words, the objective is to design the joint's input torque $\tau$ that the angular position $q$ follows the desired position $q_d$ accurately (i.e., $(q_d - q) \to 0$ as $t \to \infty$, where $t$ denotes the time) even if there are uncertainty in the dynamics, vibration disturbance $\tau_v$, and payload variation $\tau_l$.

In order to design the controller by following Lyapunov analysis, we define the term (Slotine, Li and others, 1991):

$$\eta =: \Delta \dot{q} + \sigma \Delta q \tag{3}$$

where $\eta \in \Re^n$ is a filtered signal of $\Delta q$; $\sigma \in \Re^{n \times n}$ is a positive definite diagonal gain matrix; $\Delta q$ is the position error.

**Assumption 2.** The desired link position $q_d$ is known; the current position and velocity $q$ and $\dot{q}$ are measurable due using of an encoder. This also makes $\eta$, $\Delta q$, and $\Delta \dot{q}$ measurable.

**Assumption 3.** The acceleration $\ddot{q}$ is not measurable, because measuring the acceleration requires differentiating the velocity which produces noise.

### 3.2. Analysis and Control Design

For the design purpose that fulfills our objective, we define the following Lyapunov function candidate

$$V := \frac{1}{2} \eta^T M(q) \eta + \frac{1}{2} \tilde{\phi}^T \Lambda^{-1} \tilde{\phi} \tag{4}$$

where $V \in \Re$; $\tilde{\phi}$ is subsequent term that will be defined later; $\Lambda \in \Re^{n \times n}$ is a diagonal matrix. Taking the time derivative of Eq. (4) gives

$$\dot{V} = \frac{1}{2} \dot{\eta}^T M(q) \eta + \frac{1}{2} \eta^T \dot{M}(q) \eta + \frac{1}{2} \eta^T M(q) \dot{\eta} + \frac{1}{2} \dot{\tilde{\phi}}^T \Lambda^{-1} \tilde{\phi} + \frac{1}{2} \tilde{\phi}^T \Lambda^{-1} \dot{\tilde{\phi}}. \tag{5}$$

Since all parts in the above equation are scalar quantities, we can take the transpose of $\frac{1}{2} \dot{\eta}^T M(q) \eta$ and $\frac{1}{2} \dot{\tilde{\phi}}^T \Lambda^{-1} \tilde{\phi}$ which simplifies Eq. (5) to be

$$\dot{V} = \frac{1}{2} \eta^T \dot{M}(q) \eta + \eta^T M(q) \dot{\eta} + \tilde{\phi}^T \Lambda^{-1} \dot{\tilde{\phi}}. \tag{6}$$

Utilizing the term $M(q)\dot{\eta}$ in Eq. (6), we take the derivative of (3) and multiply it by $M(q)$ to obtain

$$M(q)\dot{\eta} = M(q)\Delta \ddot{q} + \sigma M(q) \Delta \dot{q}. \tag{7}$$

Equation (7) can be extended by substituting $\Delta \ddot{q} = \ddot{q}_d - \ddot{q}$ to obtain

$$M(q)\dot{\eta} = M(q)\ddot{q}_d - M(q)\ddot{q} + \sigma M(q) \Delta \dot{q}. \tag{8}$$

Using $M(q)\ddot{q} = \tau - \tau_v - \tau_l - V_m(q, \dot{q})\dot{q} - G(q)$ from Eq. (1), we can write Eq. (8) as

$$M(q)\dot{\eta} = M(q)\ddot{q}_d - \tau + \tau_v + \tau_l + V_m(q, \dot{q})\dot{q} + G(q) + \sigma M(q)\Delta \dot{q}. \tag{9}$$

Since $\dot{q} = \dot{q}_d - \Delta \dot{q}$ and $\Delta \dot{q} = \eta - \sigma \Delta q$, we will replace $\dot{q}$ so we rewrite Eq. (9) as

$$M(q)\dot{\eta} = M(q)\ddot{q}_d - \tau + \tau_v + \tau_l$$
$$+ V_m(q,\dot{q})[\dot{q}_d - \eta + \sigma\Delta q] \quad (10)$$
$$+ G(q) + \sigma M(q)\Delta\dot{q}.$$

We define

$$Y\phi := V_m(q,\dot{q})\dot{q}_d + V_m(q,\dot{q})\sigma\Delta q$$
$$+ \sigma M(q)\Delta\dot{q} + G(q) \quad (11)$$
$$+ M(q)\ddot{q}_d$$

where $Y \in \Re^{n \times m}$ is the regression matrix that has the states of the system which are linear in the parameters; $\phi \in \Re^n$ is the parameters vector which has unknown positive constants. Substituting (11) into (10) yields

$$M(q)\dot{\eta} = Y\phi - V_m(q,\dot{q})\eta - \tau + \tau_v + \tau_l. \quad (12)$$

To design the control input $\tau$ we utilize Eq. (12) which is the open loop error system. First, we use the certainty equivalence approach to deal with the term $Y\phi$. Second, we use two gain matrices to deal with both error in the system and the disturbance torques. Since it is possible to inject measurable terms in the input torque, input torque $\tau$ is designed as following

$$\tau = Y\hat{\phi} + K_1\eta + K_2 sgn(\eta) \quad (13)$$

where $\hat{\phi}$ is the estimate of the unknown parameter in the system; $K_1$, and $K_2$ are positive diagonal gain matrices; $sgn(\eta)$ is the signum function that is defined as

$$sgn(\eta) := \begin{cases} -1, & \eta < 0 \\ 0, & \eta = 0 \\ 1, & \eta > 0. \end{cases}$$

The estimate of the unknown parameters $\hat{\phi}$ means that we do not exactly know $\phi$, we just know an approximation of it. By substituting (13) into (12) we get

$$M(q)\dot{\eta} = Y\tilde{\phi} - V_m(q,\dot{q})\eta - K_1$$
$$- K_2 sgn(\eta) + \tau_v + \tau_l \quad (14)$$

where $\tilde{\phi} = \phi - \hat{\phi}$ is the mismatch between the actual and the estimated parameters.

Now in order to investigate the validity of the designed input torque $\tau$ in (13), we substitute (14) into (6) that gives

$$\dot{V} = \frac{1}{2}\eta^T \dot{M}(q)\eta - \eta^T V_m(q,\dot{q})\eta + \eta^T[Y\tilde{\phi}$$
$$- K_1\eta - K_2 sgn(\eta) + \tau_v \quad (15)$$
$$+ \tau_l] + \tilde{\phi}^T \Lambda^{-1}\dot{\tilde{\phi}}.$$

By calling Property 1, we can write Eq. (15) as

$$\dot{V} = \eta^T[Y\tilde{\phi} - K_1\eta - K_2 sgn(\eta) + \tau_v + \tau_l]$$
$$+ \tilde{\phi}^T \Lambda^{-1}\dot{\tilde{\phi}}. \quad (16)$$

Since $\tilde{\phi} = \phi - \hat{\phi}$ and $\phi$ is constant, $\dot{\tilde{\phi}} = -\dot{\hat{\phi}}$. Therefore, we replace $\dot{\tilde{\phi}}$ with $-\dot{\hat{\phi}}$ in Eq. (16) so it is written as

$$\dot{V} = \eta^T[Y\tilde{\phi} - K_1\eta - K_2 sgn(\eta) + \tau_v + \tau_l]$$
$$- \tilde{\phi}^T \Lambda^{-1}\dot{\hat{\phi}}. \quad (17)$$

Now, we design the estimate parameter $\dot{\hat{\phi}} = \Lambda Y^T \eta$. By taking the transpose of the term $Y\eta^T\tilde{\phi}$ in (17) and substituting the designed $\dot{\hat{\phi}}$ in the same equation, it gives

$$\dot{V} = -\eta^T K_1\eta - \eta^T K_2 sgn(\eta)$$
$$+ \eta^T(\tau_v + \tau_l). \quad (18)$$

If we upper bound the norm of the disturbances $||\tau_v + \tau_l||$ by a constant $c$, and we pick the elements of $K_2$ to be bigger than the constant $c$ so it can dominant it, the derivative of the Lyapunov function $\dot{V}$ in (18) can be upper bounded as follows

$$\dot{V} \leq -K_1||\eta||^2. \quad (19)$$

The above result makes the derivative of the Lyapunov function a negative semi-definite function. Therefore, we apply theorem 8.4 in (Khalil, 2002) and conclude that $\eta \to 0$ as $t \to \infty$. Additionally, since the error $\Delta q$ is a low pass filter of $\eta$, this means that what happens to $\eta$ happens to the error $\Delta q$. Eventually, we conclude that the error $\Delta q$ approaches zero as time goes to infinity.

## 4. ROBOT MODEL FORMULATION AND SIMULATION RESULTS

In this section, the results of the simulation verification and validation of the proposed control approach are reported. Therefore, a two-link robot manipulator is formulated, two disturbance signals are generated to represent the vibration and payload variation torques, and the proposed controller is applied. The robot model, disturbances, and the proposed controller are simulated in MATLAB-SIMULINK as shown in Figure 1.

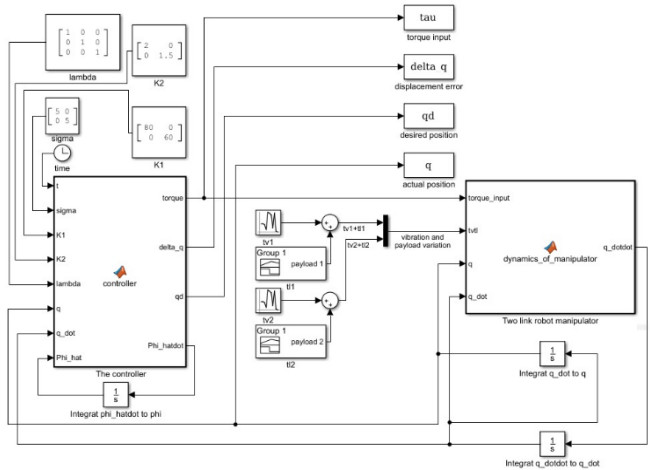

**Figure 1:** Simulation diagram of the control design and the robot dynamics

Each block contains a function that is built according to the mathematical analysis in the previous section. Table 1 presents the physical parameters of the two-link robot manipulator.

**Table (1)** Physical parameters of the two-link manipulator

| Symbol | Description | Value | Unit |
|---|---|---|---|
| $m_1$ | Mass of link 1 | 0.5 | kg |
| $m_2$ | Mass of link 2 | 0.4 | kg |
| $l_1$ | Length of link 1 | 0.6 | M |
| $l_2$ | Length of link 2 | 0.5 | M |

Each link is modelled as a uniform rectangular bar. We use Eq. (1) to represent the dynamics of the robot as follows:

$$\begin{bmatrix} M_{11} & M_{12} \\ M_{21} & M_{22} \end{bmatrix} \begin{bmatrix} \ddot{q}_1 \\ \ddot{q}_2 \end{bmatrix} + \begin{bmatrix} V_1 \\ V_2 \end{bmatrix} + \begin{bmatrix} G_1 \\ G_2 \end{bmatrix} = \begin{bmatrix} \tau_1 \\ \tau_2 \end{bmatrix} - \begin{bmatrix} \tau_{v1} \\ \tau_{v2} \end{bmatrix} - \begin{bmatrix} \tau_{l1} \\ \tau_{l2} \end{bmatrix} \quad (20)$$

Where

$$M_{11} = (m_1 + m_2)l_1^2 + m_2 l_2^2 + 2 m_2 l_1 l_2 \cos(q_1)$$
$$M_{12} = m_2 l_2^2 + m_2 l_1 l_2 \cos(q_2)$$
$$M_{21} = m_2 l_2^2 + 2 m_2 l_1 l_2 \cos(q_2)$$
$$M_{22} = m_2 l_2^2$$
$$V_1 = -m_2 l_1 l_2 (2\dot{q}_1 \dot{q}_2 + \dot{q}_2^2)\sin(q_2)$$
$$V_2 = m_2 l_1 l_2 \dot{q}_1^2 \sin(q_2)$$
$$G_1 = (m_1 + m_2)gl_1 \cos(q_1) + m_2 gl_2 \cos(q_1 + q_2)$$
$$G_2 = m_2 gl_2 \cos(q_1 + q_2)$$

and the Earth gravity constant $g = 9.807 \, m/s^2$. The disturbance torques $\tau_v$ and $\tau_l$ will be presented subsequently. All of the above equations are written in the block that is representing the dynamics of the manipulator in Figure 1.

The desired positions for both links are set as follows:

$$\begin{bmatrix} q_{d1} \\ q_{d2} \end{bmatrix} = \begin{bmatrix} 114.95° \sin(1.5t)e^{-0.03} \\ 85.94° \cos(2t)e^{-0.03} \end{bmatrix} \quad (21)$$

and their initial conditions are:

$$\begin{bmatrix} q_{d1}(0) \\ q_{d2}(0) \end{bmatrix} = \begin{bmatrix} 0 \\ 85.94° \end{bmatrix}. \quad (22)$$

The disturbance signals are formed to match the mentioned assumptions. Therefore, the disturbance torque of vibration $[\tau_{v1} \; \tau_{v2}]^T$ is composed by a bounded Gaussian noise. Therefore, we set a mean value $= [0 \; 0]^T$, a variance $= [0.01 \; 0.015]^T$, and a sampling time $0.01s$.

The disturbance torques of the payload variation for both links are represented respectively as follows:

$$\tau_{l1} = \begin{cases} 0.65 \text{ Nm}, & 4s \le t \le 8s \\ 0.15 \text{ Nm}, & 8s \le t \le 10s \\ 0, & \text{otherwise} \end{cases}$$

$$\tau_{l2} = \begin{cases} 0.75 \text{ Nm}, & 4s \le t \le 8s \\ 0.25 \text{ Nm}, & 8s \le t \le 10s \\ 0, & \text{otherwise.} \end{cases}$$

These signals are represented separately as shown in block diagram in Figure 2.

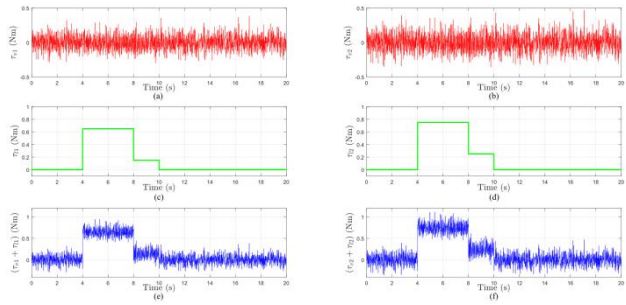

**Figure 2:** Disturbance torques due to vibration and payload variation: (a) and (b) vibration effect; (c) and (d) payload variation effect; (e) and (f) effect of combination on link 1 and link 2, respectively.

Also, the disturbances' effect on the first and second links is shown in the same figure. The bounds of these disturbances are:

$$\begin{bmatrix} -0.2873 \\ -0.3518 \end{bmatrix} \leq \begin{bmatrix} \tau_{v1} + \tau_{l1} \\ \tau_{v2} + \tau_{l2} \end{bmatrix} \leq \begin{bmatrix} 0.9446 \\ 1.1108 \end{bmatrix}$$

The controller, that is derived analytically, and the desired trajectories are written in the controller block in Figure 1. All control gains are chosen to yield the best performance using trial and error method. Therefore, the control gains for the proposed control design are selected as $K_1 = diag[80 \ 60]$, $K_2 = diag[2 \ 1.5]$, $\sigma = diag[5 \ 5]$, and $\Lambda = diag[1 \ 1 \ 1]$. The input torques of the proposed controller is shown in Figure 3.

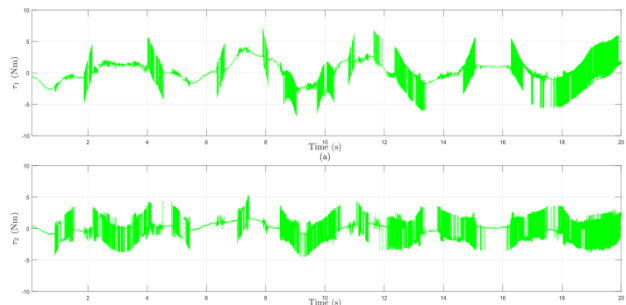

**Figure 3:** Torque inputs of the proposed controller: (a) joint 1 (b) joint 2.

The input torque values for both joints have high oscillations with respect to time, however, these oscillations are not beyond the classical bandwidth limit of the actuators which are driving the robot links. However, the law values of input torque are relatively law compared with other approaches from the literature. Therefore, these input torque can be implemented at low cost and they do not saturate the actuators which are driving the robot links.

The proposed controller performance is shown in Figure 4.

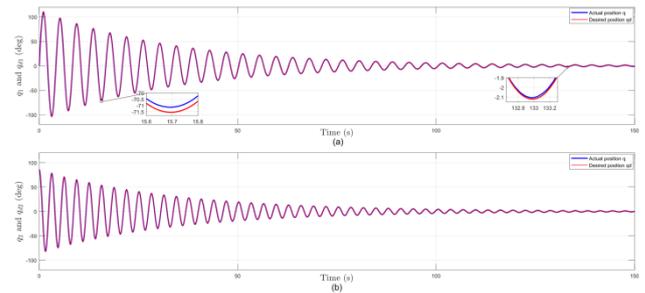

**Figure 4:** Tracking performance of the proposed control approach: (a) and (b) desired and actual position of link 1 and link 2, respectively;

This figure depicts the robot actual angular displacements $q_1$ and $q_2$ with respect to time, and their desired references $q_{d1}$ and $q_{d2}$, respectively. We notice that the errors $\Delta q_1$ and $\Delta q_2$ approach zero as time passes as shown in Figure 5.

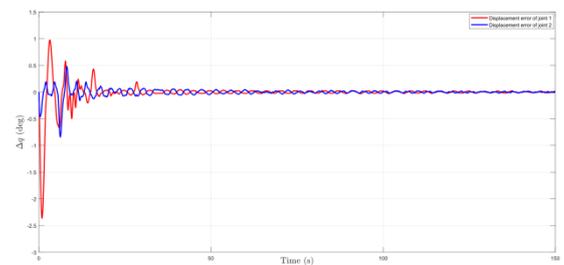

**Figure 5:** Angular displacement error of link 1 and link 2.

To provide further explanation and validate the performance of the proposed controller, we use the following statistical indices for the angular displacement error and the input torque:

1. Maximum absolute values of the displacement error for each link.

$$\Delta q_{imax} = \max_{j=1,\dots,M}(|\Delta q_i(j)|) \quad (23)$$

2. Root mean square rms values of the displacement error and input torque for each link and joint, respectively.

$$\Delta q_{irms} = \sqrt{\frac{1}{M}\sum_{j=1}^{M}||\Delta q_i(j)||^2} \quad (24)$$

$$\tau_{irms} = \sqrt{\frac{1}{M}\sum_{j=1}^{M}||\tau_i(j)||^2} \quad (25)$$

where $i$ is the link/joint number and $M$ is the number of sampling steps of the simulation. Table 2 presents the indices in (23), (24), and (25) for both links/joints obtained from Figure 3 and Figure 4.

**Table (2)** Performance summary for the proposed controller

| Indexes | First link/joint | Second link/joint |
|---|---|---|
| $\Delta q_{max}(\text{deg})$ | 0.9756 | 0.4804 |
| $\Delta q_{irms}(\text{deg})$ | 0.1199 | 0.0385 |
| $\tau_{irms}(\text{Nm})$ | 1.8105 | 1.6727 |

According to the physical parameters of the robot in Table 1, the external disturbances, and the obtained results in Table 2, we see that the proposed controller is implementable, efficient, and accurate. Moreover, the time for evaluating control actions of the proposed controller is presented in Table 3.

**Table (3)** Performance of the proposed controller in terms of computation time

| No. of calls | Time/Call (ms) | Total time (s) |
|---|---|---|
| 344013 | 0.0062 | 2.14 |

The oscillation of the input torques with settling time for both links are presented in Table 4.

**Table (4)** Oscillation and settling time of the input torques

| Link/joint order | Maximum oscillation (kH$_z$) | Settling time (s) |
|---|---|---|
| First | 1.01 | 15.2 |
| Second | 0.98 | 13.5 |

As presented, the settling time is high; however, the amplitude of the error is not significant; therefore, it does not affect the performance and stability of the system. Furthermore, the input torques oscillation are not beyond the classical bandwidth limits of the actuators.

## 5. Conclusion

In this paper, we present an adaptive-sliding mode approach for position control of robot manipulators, in joint space, in the presence of uncertain torques due to vibration and payload variation. Firstly, the problem of tracking control for an n-degree-of-freedom, revolute-joints, serial robot arm is formulated. Secondly, the controller is designed based on the Lyapunov analysis and following the adaptive and sliding mode control approach. The rms values of the error are 0.119 deg and 0.03 deg for link 1 and link 2, respectively. Which means, despite the uncertainty that is caused by the uncertain dynamics, vibration, and payload variation, the controller is shown to guarantee asymptotic tracking displacement errors. Additionally, the input torque rms values are 1.81 Nm and 1.67 Nm for joint 1 and joint 2, respectively. These values are relatively law compared with other approaches from the literature despite the high oscillation. This makes the controller implementable at low cost and reasonable control effort.

## Acknowledgements

This study was supported by Salahaddin University-Erbil, Iraq, and the University of Florida, FL, USA, in the Framework of a split-side Ph.D. program.

## Conflict of Interest

The authors declare no conflict of interest.